%% file: conference_101719.tex
\documentclass[Journal]{IEEEtran}
\IEEEoverridecommandlockouts
\usepackage{cite}
\usepackage{amsmath,amssymb,amsfonts}
\usepackage{hyperref}
\usepackage{algorithmic}
\usepackage{graphicx}
\usepackage{textcomp}
\usepackage{fancyhdr}
\usepackage{float}
\usepackage[table]{xcolor} 
\definecolor{lightgray}{rgb}{0.83, 0.83, 0.83}
\usepackage{xcolor}
\usepackage{tcolorbox}
\usepackage{ulem}
\usepackage{tcolorbox}
\usepackage{adjustbox}
\usepackage{array}
\usepackage{fontawesome5}
\def\BibTeX{{\rm B\kern-.05em{\sc i\kern-.025em b}\kern-.08em
    T\kern-.1667em\lower.7ex\hbox{E}\kern-.125emX}}
    
\begin{document}

\title{Ethical Aspects of ChatGPT in Software Engineering Research\\
%{\footnotesize \textsuperscript{*}Note: Sub-titles are not captured in Xplore and should not be used}

}

\author{
    \IEEEauthorblockN{
        Muhammad Azeem Akbar\IEEEauthorrefmark{1},
        Arif Ali Khan\IEEEauthorrefmark{2},
        Peng Liang\IEEEauthorrefmark{3}
    }
    
    \IEEEauthorblockA{\IEEEauthorrefmark{1}\textit{Software Engineering Department, Lappeenranta-Lahti University of Technology}, 15210 Lappeenranta, Finland}
    
    \IEEEauthorblockA{\IEEEauthorrefmark{2}\textit{M3S Empirical Software Engineering Research Unit, University of Oulu}, 90014 Oulu, Finland}
    
    \IEEEauthorblockA{\IEEEauthorrefmark{3}\textit{School of Computer Science, Wuhan University}, 430072 Wuhan, China}
    
    \IEEEauthorblockA{\IEEEauthorrefmark{1}azeem.akbar@ymail.com, \IEEEauthorrefmark{2}arif.khan@oulu.fi, \IEEEauthorrefmark{3}liangp@whu.edu.cn}
}

\maketitle

\begin{abstract}
ChatGPT can improve Software Engineering (SE) research practices by offering efficient, accessible information analysis and synthesis based on natural language interactions. However, ChatGPT could bring ethical challenges, encompassing plagiarism, privacy, data security, and the risk of generating biased or potentially detrimental data. This research aims to fill the given gap by elaborating on the key elements: motivators, demotivators, and ethical principles of using ChatGPT in SE research. To achieve this objective, we conducted a literature survey, identified the mentioned elements, and presented their relationships by developing a taxonomy. Further, the identified literature-based elements (motivators, demotivators, and ethical principles) were empirically evaluated by conducting a comprehensive questionnaire-based survey involving SE researchers. Additionally, we employed Interpretive Structure Modeling (ISM) approach to analyze the relationships between the ethical principles of using ChatGPT in SE research and develop a level based decision model. We further conducted a Cross-Impact Matrix Multiplication Applied to Classification (MICMAC) analysis to create a cluster-based decision model. These models aim to help SE researchers devise effective strategies for ethically integrating ChatGPT into SE research by following the identified principles through adopting the motivators and addressing the demotivators. The findings of this study will establish a benchmark for incorporating ChatGPT services in SE research with an emphasis on ethical considerations.

{\textit{Impact Statement—}}This paper establishes the impact of employing ChatGPT in SE research while carefully addressing ethical challenges like privacy, data security, and bias. By developing a taxonomy through a comprehensive literature survey and an extensive questionnaire-based survey, we have created a guideline based on the identified motivators, demotivators, and ethical principles. Using ISM and MICMAC approaches, we have developed decision models to help strategize the ethical integration of ChatGPT into SE research. This pioneering study creates a benchmark for incorporating AI services in SE research, emphasizing the balance of harnessing the potential benefits while mitigating ethical risks.

\end{abstract}


\input{introduction}
\input{stage}
\input{results}

\input{Study-Implications}

\input{Threats-to-Validity}
\input{Conclusions-Future-Plan}
\bibliographystyle{IEEEtran}
\bibliography{conference_101719}

\end{document}

%% file: introduction.tex
\section{Introduction}
\label{sec:introduction}

ChatGPT is a cutting-edge language model created by OpenAI \cite{brown2020language}, designed to generate human-like responses to various prompts. The model employs deep learning algorithms, utilizing the latest techniques in Natural Language Processing (NLP) to generate relevant and coherent responses. GPT, or ``Generative Pre-trained Transformer'' refers to the model's architecture based on the transformer architecture and pre-trained on a vast corpus of textual data \cite{sohail2023future}. ChatGPT has been fine-tuned on conversational data, allowing it to generate appropriate and engaging responses in a dialogue context \cite{brown2020language, vaswani2017attention}. The model's versatility means that it can be applied to numerous applications, including chatbots, virtual assistants, customer service, and automated content creation. The OpenAI team continues to update and improve the model with the latest data and training techniques, ensuring it remains at the forefront of NLP research and development \cite{OpenAI}.

ChatGPT has significant potential for use in academic research \cite{bukar4381394text}, particularly for performing SE activities \cite{white2023chatgpt}. Researchers can utilize ChatGPT to generate realistic and high-quality text for various applications, including language generation, language understanding, dialogue systems, and experts' opinion transcripts \cite{perkins2023academic}. ChatGPT can also be fine-tuned for specific domains or tasks, making it a flexible tool for researchers to create customized language models \cite{dwivedi2023so}. In addition, ChatGPT can be used to generate synthetic data for training other models, and its performance can be evaluated against human-generated data. Moreover, ChatGPT can be used for research on social and cultural phenomena related to language use. For example, researchers can use ChatGPT to simulate conversations and interactions between people with different cultural backgrounds or to investigate the impact of linguistic factors such as dialect, jargon, or slang on language understanding and generation \cite{neumann2023we}.

ChatGPT significantly impacts research, particularly in qualitative research using NLP tools. Its ability to generate high-quality responses has made it a valuable tool for language generation, understanding, and dialogue systems \cite{van2023chatgpt}. Researchers can leverage ChatGPT to save time and resources, create customized language models, and fine-tune for specific domains or tasks \cite{van2023chatgpt}. ChatGPT's simulation capabilities also allow researchers to understand natural language in different contexts and develop more nuanced language models \cite{neumann2023we, mhlanga2023open}. Overall, ChatGPT has advanced the field of NLP and paved the way for more advanced language models and applications \cite{du2023chat}.

ChatGPT behaves as a smart, intelligent, and effective tool for SE research \cite{m2022exploring, bishop2023computer, lundchatgpt}. For instance, the ChatGPT can be used in literature review-based research to extract data by giving specific queries and related text in quotes. Similarly, we noticed that the ChatGPT is also an effective tool for generating the codes, concepts, and categories from transcripts in qualitative research \cite{mesec2023language}. Considering the effectiveness and usability of ChatGPT in academic research, we conducted this study (1) to explore and understand the motivators (positive factors) and demotivators (negatively influencing factors) across the ethical aspects (principles) of ChatGPT in SE research and (2) to develop Interpretive Structure Modeling (ISM) and Cross-Impact Matrix Multiplication Applied to Classification (MICMAC) based decision-making models in order to understand the relationships between ethical principles for using ChatGPT in SE research. We believe that the outcomes of this research will benefit the academic research community by providing a body of knowledge and serving as guidelines for considering ChatGPT in SE research.

The rest of the papers is organized as follows - the research methodology is presented in Section \ref{sec:stage}, the results are discussed in Section \ref{sec: Results and Discussions} and the implications of the study findings are reported in Section \ref{Sec: Study Implications}. The threats to the validity of the study findings are highlighted in Section \ref{Sec:Threats to Validity} and finally, we concluded the study with future avenues in Section \ref{Sec:Conclusions and Future Plans}.

%% file: stage.tex
\section{Setting the stage}
\label{sec:stage}
The research aims to comprehensively understand the ethical implications and potential threats associated with using ChatGPT in SE research and develop guidelines and recommendations for responsible research practices to mitigate these issues and threats \cite{sallam2023utility}. By promoting the responsible and ethical use of ChatGPT \cite{m2022exploring}, our research aims to help the SE research community benefit from the important motivators, demotivators, and ethical principles of using ChatGPT in SE research. 

In order to establish the objective of this study, we began with a literature survey to examine the factors that motivate and demotivate SE researchers, as well as the ethical principles of using ChatGPT in SE research. Subsequently, we sought validation of our literature findings by engaging expert researchers through a questionnaire survey study. Finally, we employed Interpretive Structure Modeling (ISM) to develop the decision-making model based on the complex relationship between the ethics principles of using ChatGPT in SE research. A visual representation of our methodology can be found in Figure \ref{Fig.Proposed research method}, with a concise discussion of each step provided in the following sections.

\subsection{Literature Survey} \label{Sec: literature Survey}
To identify the motivators, demotivators, and principles associated with the ethical use of ChatGPT in SE research, we conducted a literature survey, examining both peer-reviewed published articles and grey literature\cite{akbar2020requirements, akbar2021fuzzy}. Using the common keywords, we explored the grey literature across general Google search and Google Scholar to investigate peer-reviewed literature studies. Furthermore, we employed the snowballing data sampling approach to collect potential literature material related to the study objective \cite{wohlin2014guidelines}. This involved examining reference sections of selected studies (backward snowballing) and citations (forward snowballing), resulting in increasing the sample size by including more relevant studies \cite{wohlin2014guidelines}.

\subsection{Questionnaire Survey Study} \label{Sec: Questionnaier Survey}
The questionnaire survey is an appropriate approach to collect the data from a large and targeted population \cite{patten2016questionnaire}. In this study, we designed a survey questionnaire to validate the identified motivators, demotivators, and principles for evaluating the ethical implications of ChatGPT in SE research. We divided the questionnaire into two parts. The first part focuses on the demographics of survey participants, while the second part consists of the identified motivators, demotivators, and principles. We used the five-point Likert scale \textit{(strongly agree, agree, neutral, disagree, and strongly disagree)} to encapsulate the opinions of the targeted population. The second part of the questionnaire also includes an open-ended question, enabling participants to suggest any additional motivators, demotivators, or principles overlooked during the literature survey. 

To reach the target population, we developed an online questionnaire using Google Forms and sent invitations via personal email, organizational email, and LinkedIn. We employed the snowball sampling approach to collect a representative data sample by encouraging participants to share the questionnaire across their research network. Snowball sampling is efficient, cost-effective, and suitable for large, dispersed target populations \cite{wohlin2014guidelines}. Data collection took place from 15 January to 25 April 2023, returning 121 responses, of which 113 were used for further analysis after removing eight incomplete responses.

We used the frequency analysis approach to analyze the collected data, which is appropriate for the descriptive type of data analysis \cite{von2003configural}. This approach compares survey variables and computes the agreement level among participants based on the selected Likert scale. Frequency analysis has also been used in other software engineering studies \cite{niazi2016toward,khan2021agile}.

Furthermore, in order to address the ethical considerations pertaining to our survey respondents and the gathered data, we incorporated a detailed consent form at the outset of the questionnaire. This form thoroughly addresses all relevant aspects of data privacy and the confidentiality of participant identities. Every respondent must agree to the terms and conditions of the survey questionnaire prior to offering their feedback. A sample of the survey questionnaire can be found at the link\footnote{\url{https://tinyurl.com/pwbr7mzw}}.

\subsection{ISM Approach} \label{Sec:ISM Approach}
\begin{figure}[htbp]
\centerline{\includegraphics[width=\linewidth]{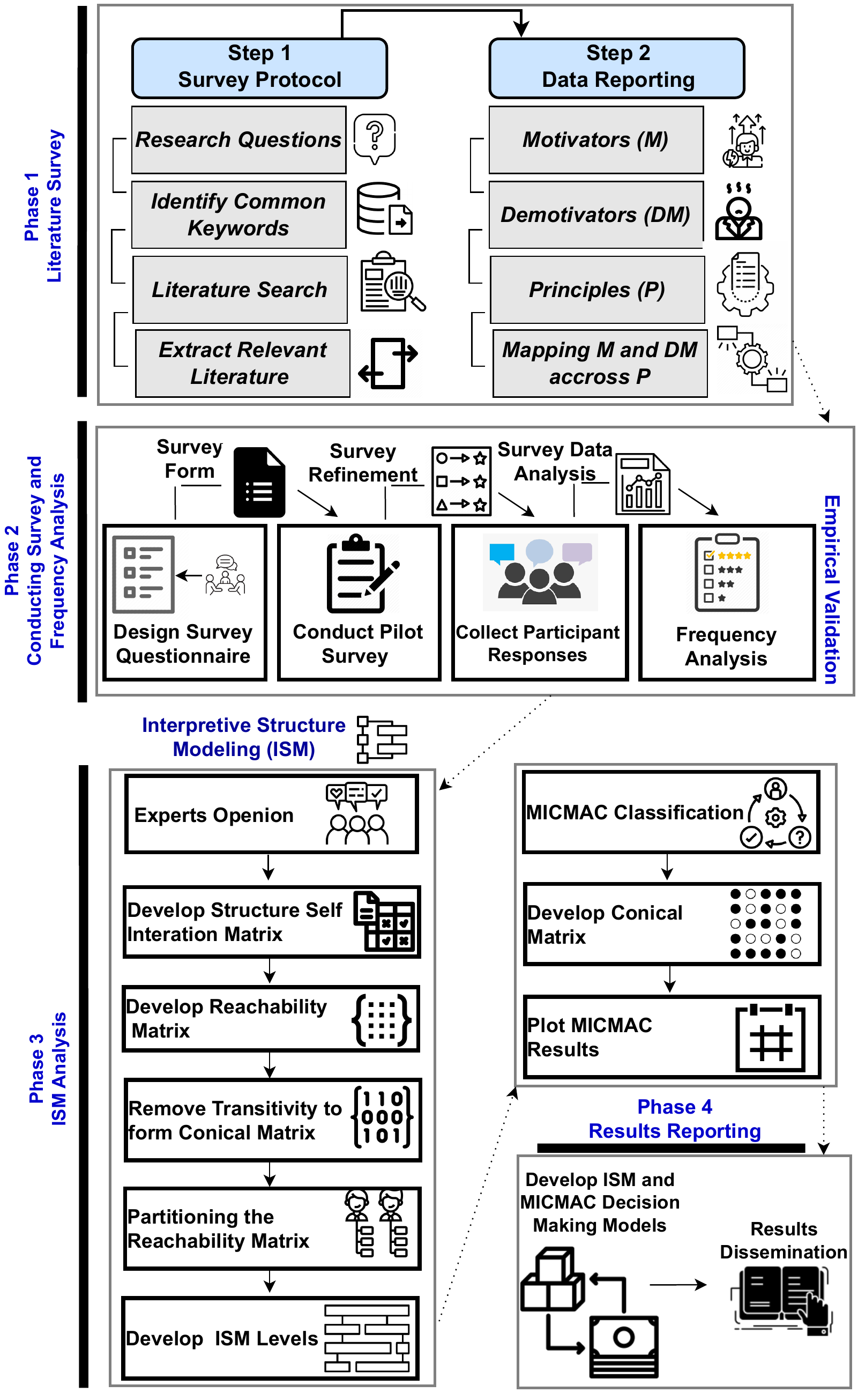}}
\caption{Proposed Research Methodology}
\label{Fig.Proposed research method}
\end{figure}

Interpretive Structure Modeling (ISM) is an interactive learning process approach introduced by Forrester \cite{forrester1961industrial}, that establishes a map of complex relationships among various factors, resulting in a comprehensive system model. The model provides a clear and conceptual representation in a graphical format \cite{ravi2005analysis}. ISM simplifies the complexities associated with relationships among different aspects, offering a better understanding of such factor relationships. Several relevant software engineering studies have employed this approach to develop conceptual models that clarify the relationships between principles \cite{kannan2009hybrid, sharma1995objectives, agarwal2017modeling}. We used the ISM approach to identify the interactions between the ethical principles (variables) of adopting ChatGPT in SE research. Figure \ref{Fig.Proposed research method} illustrates the detailed steps involved in the ISM approach and elaborated as follows \cite{malone1975introduction}.

\begin{itemize}
    \item Define a contextual relationship that captures the dependencies among the research ethics principles. For example, “Principal A influences Principal B".
    \item Create an initial reachability matrix that captures the direct relationships between the principles based on the contextual relationship defined in Step 1.
    \item Compute the final reachability matrix by considering both direct and indirect relationships among the principles.
    \item Using the final reachability matrix, partition the principles into different levels based on their relationships, starting from the least influential principles at the bottom level and moving up to the most.
\end{itemize}

The ISM approach was applied by inviting respondents from the first survey to participate in an ISM decision-making survey - 23 experts agreed to join. Their insights were collected using a seperate questionnaire, a sample of which is provided at the given link\footnote{\url{https://tinyurl.com/5b9c4hjr}}. The collected data were then used to develop the Structural Self-Interaction Matrix (SSIM) matrix, although the sample size could potentially limit the generalizability of the study. Nonetheless, previous research has shown that studies with as few as five experts can be effective for similar decision-making processes. 

For instance, Kannan et al. \cite{kannan2009hybrid}  used the opinions of five experts for the selection of reverse logistic providers. Similarly, Soni et al. \cite{soni2015end} established a nine-member group to analyze factors contributing to complexities in an urban rail transit system. Furthermore, Attri et al. \cite{attri2013analysis}  used the input from five experts to make a decision regarding success factors for total productive maintenance. Thus, in light of existing research, we concluded that the sample of twenty-three experts in our study was adequate for the ISM-based analysis.

%% file: results.tex
\section{Results and Discussions}
\label{sec: Results and Discussions}

In this section, we provide the results and discussions, which are based on the mutual agreement of all authors. Section \ref{Sec:Literature Survey Findings} presents the results of the literature survey - defining identified motivators, demotivators, and ethical principles of using ChatGPT in SE research. In Section \ref{Sec:Empirical Study Findings}, the participants' perceptions regarding the identified motivators, demotivators, and ethical principles are discussed. Finally, Section \ref{Sec:ISMA Based Modeling of Ethical Principles} details the results of an ISM-based analysis of the identified principles. 

\subsection{Literature Survey Findings} \label{Sec:Literature Survey Findings}
This section presents the findings derived from both grey literature and peer-reviewed studies. The study reveals the significant motivators, demotivators, and their association with relevant ethical principles of using ChatGPT in SE research (see the subsequent sections).

\subsubsection{Motivators of using ChatGPT in SE research}
ChatGPT offers a valuable tool for SE research in several ways. Firstly, ChatGPT can generate synthetic data for software testing, which is essential in SE \cite{jalil2023chatgpt}. This can save time and resources by automating the process of generating test cases, allowing for rapid iteration and software performance evaluation \cite{jalil2023chatgpt}. ChatGPT can be fine-tuned for specific SE domains, such as requirements engineering or software quality assurance \cite{du2023chat}. This can help researchers to create customized language models that can be used to study different facets of SE. 

Thirdly, ChatGPT can be used to simulate user interactions with software systems, allowing researchers to test and evaluate software usability and user experience. By simulating user interactions, researchers can identify potential issues and improve the overall design and functionality of the software system \cite{jalil2023chatgpt}. Therefore, by leveraging ChatGPT, the researchers can elevate the level of SE research and develop more sophisticated and effective software systems. Eventually, based on the unique characteristics of ChatGPT, we have identified 14 key motivators from the existing literature that are essential to consider when utilizing ChatGPT in SE research \cite{dwivedi2023so, van2023chatgpt, lundchatgpt, ChatGPT_for_education, skjuve4376834people, frieder2023mathematical, oviedo2023risks, donmez2023conducting, qin2023chatgpt, mitrovic2023chatgpt, rahaman2023can}. Motivators are factors or features that encourage or drive a person or organization to take a particular action or make a decision. In the context of SE research, motivators can refer to the benefits or advantages of using ChatGPT, to achieve specific research goals. The identified 14 motivators are briefly elaborated as follows:

\textit{M1 (Synthetic data generation):} ChatGPT can generate synthetic data for software testing, which can save time and resources in SE research.

\textit{M2 (Domain-specific fine-tuning):} ChatGPT can be fine-tuned for specific SE domains, such as requirements engineering and software quality assurance.

\textit{M3 (Usability simulation evaluation):} ChatGPT can simulate user interactions with software systems, allowing researchers to test and evaluate software usability and user experience.

\textit{M4 (Generate requirements description):} ChatGPT can generate natural language descriptions of software requirements, making it easier for stakeholders to understand the software system.

\textit{M5 (Documentation generation improvement):} ChatGPT can be used to generate code comments and documentation, which can improve software quality and maintainability.

\textit{M6 (Bug reporting assistance):} ChatGPT can assist in software bug reporting by generating high-quality, natural language bug descriptions.

\textit{M7 (Test case generation):} ChatGPT can be used to generate test cases, enabling researchers to evaluate software performance and identify potential issues.

\textit{M8 (Automated code generation):} ChatGPT can help in automated software code generation, making it easier to build software systems and reducing the potential for human errors.

\textit{M9 (Summarize code):} ChatGPT can generate natural language summaries of software code, making it easier for developers to understand the codebase.

\textit{M10 (Maintenance assistance):} ChatGPT can help in software maintenance by generating high-quality documentation and code comments, making it easier for developers to maintain and update the software system.

\textit{M11 (Performance explanation):} ChatGPT can generate natural language explanations of software performance issues, making it easier for developers to diagnose and fix software bugs.

\textit{M12 (Generate automated report):} ChatGPT can be used to generate automated software reports, providing stakeholders with up-to-date information on software performance and quality.

\textit{M13 (Testing assistance):} ChatGPT can assist in software testing by generating test scenarios and test data, enabling researchers to evaluate software functionality and performance.

\textit{M14 (Develop user manual):} ChatGPT can generate natural language user manuals and documentation, making it easier for end-users to understand and use the software system.

Thus, ChatGPT provides a powerful and flexible tool for SE research, offering numerous motivators for researchers to incorporate it into their projects. By leveraging the power of ChatGPT, researchers can advance state-of-the-art research in SE and create more sophisticated and effective software systems.

\subsubsection{Demotivators of using ChatGPT in SE research}
While ChatGPT has several motivators for its use in SE research, there are also some demotivators to consider. ChatGPT may not always generate accurate or relevant responses, requires significant training data, and may not be suitable for certain SE tasks \cite{surameery2023use}. Its responses can be repetitive, too complex, too simple for the intended audience, and may not align with industry or domain-specific conventions. ChatGPT's responses may also reflect bias in the training data and require manual editing or correction, reducing the efficiency gains provided by automation and natural language processing \cite{van2023chatgpt, salvagno2023can}. Therefore, after reviewing the relevant literature studies \cite{van2023chatgpt, ChatGPT_for_education, donmez2023conducting, thorp2023chatgpt, sciencespo, stokelchatgpt, crawford2023leadership, cheating-students, 353wsj6d, atlas2023chatgpt, tate2023educational, ma2023abstract}, we uncovered the following demotivators (contextually, factors that can potentially limit the effectiveness of utilizing ChatGPT), which must be taken into account when using ChatGPT in SE research. 

\textit{DM1 (Model limitations acknowledged):} ChatGPT is not a perfect language model, and its responses may not always be accurate or relevant to the task at hand.

\textit{DM2 (Data-intensive fine-tuning):} ChatGPT requires a significant amount of training data to fine-tune it for specific SE tasks, which can be time-consuming and resource-intensive.

\textit{DM3 (Limited task scope):} ChatGPT may not be suitable for specific SE tasks requiring specialized knowledge or expertise outside natural language processing.

\textit{DM4 (Repetitive response issue):} ChatGPT's responses can be repetitive, which may not provide sufficient variety in generated data.

\textit{DM5 (Response complexity mismatch):} ChatGPT may generate responses that are too complex or too simple for the intended audience, making it difficult to communicate with stakeholders or end-users.

\textit{DM6 (Convention misalignment issue):} ChatGPT's responses may not always align with industry or domain-specific conventions, leading to inconsistencies and inaccuracies in generated data.

\textit{DM7 (Bias reflection issue):} ChatGPT may generate responses that are biased or reflect the bias in the training data, leading to ethical concerns and potential negative impacts on software development.

\textit{DM8 (Multilingual limitations identified):} ChatGPT is able to generate responses only in 50 languages \cite{Listoflanguages}, which may limit its usability in international software development projects.

\textit{DM9 (Integration challenges anticipated):} ChatGPT may generate responses in a format or structure incompatible with certain software development tools or platforms used in an organization's existing workflows. This incompatibility may lead to technical difficulties in integrating ChatGPT into the development process, resulting in delays and additional costs. For example, suppose ChatGPT generates code snippets in a programming language that is not supported by the development platform. In that case, developers may need to convert the code to the correct format manually. As a result, careful consideration and testing are necessary when integrating ChatGPT into an organization's software development workflow.

\textit{DM10 (Misalignment Conflicts):} ChatGPT's responses may not always match the preferences or expectations of stakeholders involved in a software development project. For example, a stakeholder may have a particular vision for a software application user interface or functionality, but ChatGPT's responses may suggest something different. This misalignment may lead to disagreements and conflicts among the project team and stakeholders, impacting the project's progress and success. Project teams need to consider the input of all stakeholders and carefully evaluate ChatGPT's suggestions to ensure they align with the project's goals and objectives. Additionally, stakeholders should be educated on the capabilities and limitations of ChatGPT to manage their expectations and ensure they are not relying solely on the tool for decision-making.

\textit{DM11 (Unrealistic responses):} ChatGPT's responses may not always align with the technical constraints of the software development environment, leading to unrealistic or impractical suggestions or recommendations.

\textit{DM12 (Demand manual editing):} ChatGPT's responses may require significant manual editing or correction, reducing the efficiency gains provided by automation and natural language processing.

Ultimately, the limitations of ChatGPT in SE research include its dependence on large amounts of training data, potential inaccuracies and biases in generated responses, limitations in language and compatibility with specific tools and platforms, potential conflicts with stakeholder preferences, and the need for significant manual editing or correction.

\subsubsection{Ethical principles of ChatGPT in SE research}
The use of ChatGPT in SE research raises numerous ethical concerns. These include issues related to bias, privacy, accountability, reliability, intellectual property, security, manipulation, unintended consequences, human labor displacement, legal compliance, ethical governance, trust, informed consent, fairness, transparency, long-term consequences, exacerbating inequalities, lack of accountability, and the ethical implications of automation \cite{dwivedi2023so, lundchatgpt, zhuo2023exploring}. Researchers have a social responsibility to ensure that ChatGPT is used ethically and in the best interests of society, with appropriate ethical governance, transparency, and accountability. Using the existing studies \cite{van2023chatgpt, mhlanga2023open, lundchatgpt, oviedo2023risks, salvagno2023can, sciencespo, stokelchatgpt, crawford2023leadership, cheating-students, tate2023educational, borji2023categorical, qadir2022engineering}, we draw the following ethical aspects (principles) of using ChatGPT in SE research.

\textit{P1 (Bias):} ChatGPT's responses may reflect the biases present in the training data, which can perpetuate existing biases and lead to unfair or discriminatory outcomes.

\textit{P2 (Privacy):} ChatGPT may generate responses that contain sensitive or personally identifiable information, potentially violating individuals' privacy rights.

\textit{P3 (Accountability):} ChatGPT's responses may not always be transparent or explainable, making it difficult to determine who is responsible for errors or biases in generated data.

\textit{P4 (Reliability):} ChatGPT may generate inaccurate or misleading responses, potentially negatively impacting software development or end-users.

\textit{P5 (Intellectual property):} ChatGPT may generate responses that infringe upon intellectual property rights, such as copyright or patent law.

\textit{P6 (Security):} ChatGPT's responses may contain sensitive information that could be exploited by malicious actors, leading to potential security breaches or cyberattacks.

\textit{P7 (Manipulation):} ChatGPT may be used to generate fake news, propaganda, or other forms of misinformation, leading to potential harm to individuals or society as a whole.

\textit{P8 (Legal compliance):} ChatGPT's responses may violate legal and regulatory requirements, such as data protection laws or accessibility standards.

\textit{P9 (Ethical governance):} The use of ChatGPT in SE research requires appropriate ethical governance, including informed consent, privacy protection, and transparency.

\textit{P10 (Trust):} ChatGPT's responses may erode trust in software development and technology more broadly, potentially leading to negative impacts on society as a whole.

\textit{P11 (Ethical decision-making):} The use of ChatGPT requires ethical decision-making and a commitment to ethical values, such as fairness, accountability, and transparency.

\textit{P12 (Social responsibility):} Researchers using ChatGPT have a social responsibility to ensure that the technology is used ethically and in the best interests of society.

\textit{P13 (Informed consent):} ChatGPT's use in SE research requires informed consent from participants, including clear explanations of the technology's risks and benefits.

\textit{P14 (Fairness):} ChatGPT must ensure that all stakeholders, including end-users, developers, and other members of the software development teams, are treated fairly.

\textit{P15 (Transparency):} ChatGPT must be transparent, with clear explanations of how the technology works, how it is being used, and what data is being collected.

\textit{P16 (Long-term consequences):} Researchers must consider the long-term consequences of using ChatGPT in SE research, including potential impacts on society, the environment, and future generations.

\textit{P17 (Ethical implications of automation):} The use of ChatGPT raises ethical questions about the implications of automation in SE, including the displacement of human labor and technical tools. 

\begin{figure*}[htbp]
\centerline{\includegraphics[width=\linewidth]{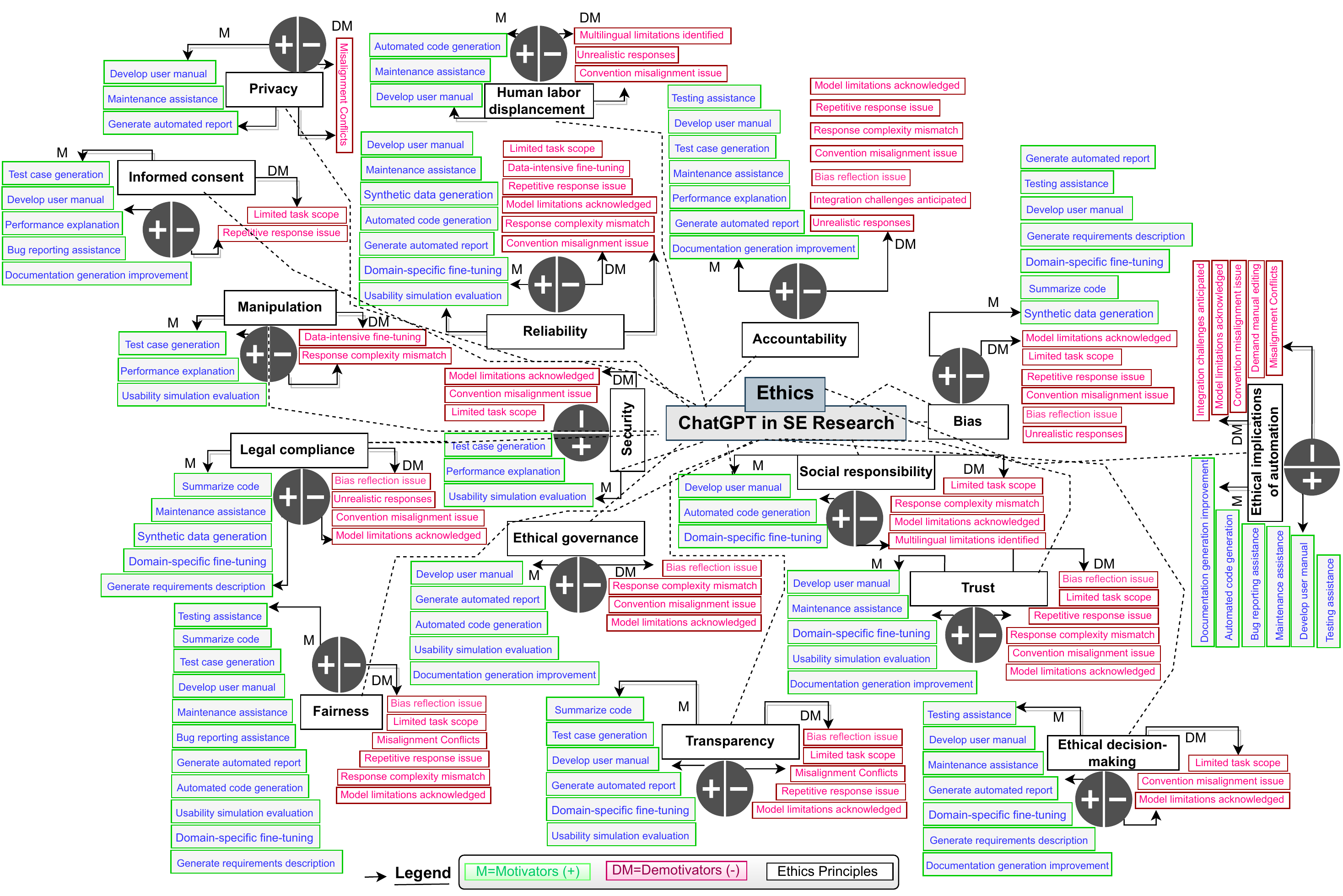}}
\caption{Mapping of motivators and demotivators against research ethics}
\label{Fig.Mapping of motivators and demotivators against research ethics}
\end{figure*}

\subsubsection{Relationship of Motivators and Demotivators with Ethical Principles}
\label{sec:relationship}

Preliminary, we develop a taxonomy by considering the identified motivators, demotivators, and their possible impacts on ChatGPT ethical principles. Motivators are factors that encourage or inspire researchers to consider ChatGPT as a tool for people to take certain actions. At the same time, demotivators discourage or hinder people from taking certain actions. In this context, motivators and demotivators may refer to factors that influence the use of ChatGPT in SE research. The taxonomy considers the possible impact of these motivators and demotivators on the ethical aspects of using ChatGPT in SE research. Ethics aspects (principles) refer to the moral standards that guide the behavior and decision-making in SE research. The proposed taxonomy (see Figure \ref{Fig.Mapping of motivators and demotivators against research ethics}) provides a roadmap for academic researchers to evaluate both the motivators (positive factors) and demotivators (negative factors) related to the ethical aspects of using ChatGPT in SE research. By considering these factors, researchers can gain a comprehensive understanding of the ethical considerations associated with using ChatGPT in SE research. This taxonomy can serve as a valuable tool for researchers to ensure that they use ChatGPT ethically and responsibly. It can also contribute to developing ethical guidelines and best practices for using ChatGPT in SE research.

\subsection{Empirical Study Findings} \label{Sec:Empirical Study Findings}
The results of the questionnaire survey study are presented in this section. Specifically, we cover (i) the demographic details of survey participants and (ii) survey participants' perceptions of motivators, demotivators, and ethical principles of using ChatGPT in SE research.

\subsubsection{Demographic Details}
\begin{figure*}[h!]
 \centering
    \includegraphics[scale=0.5]{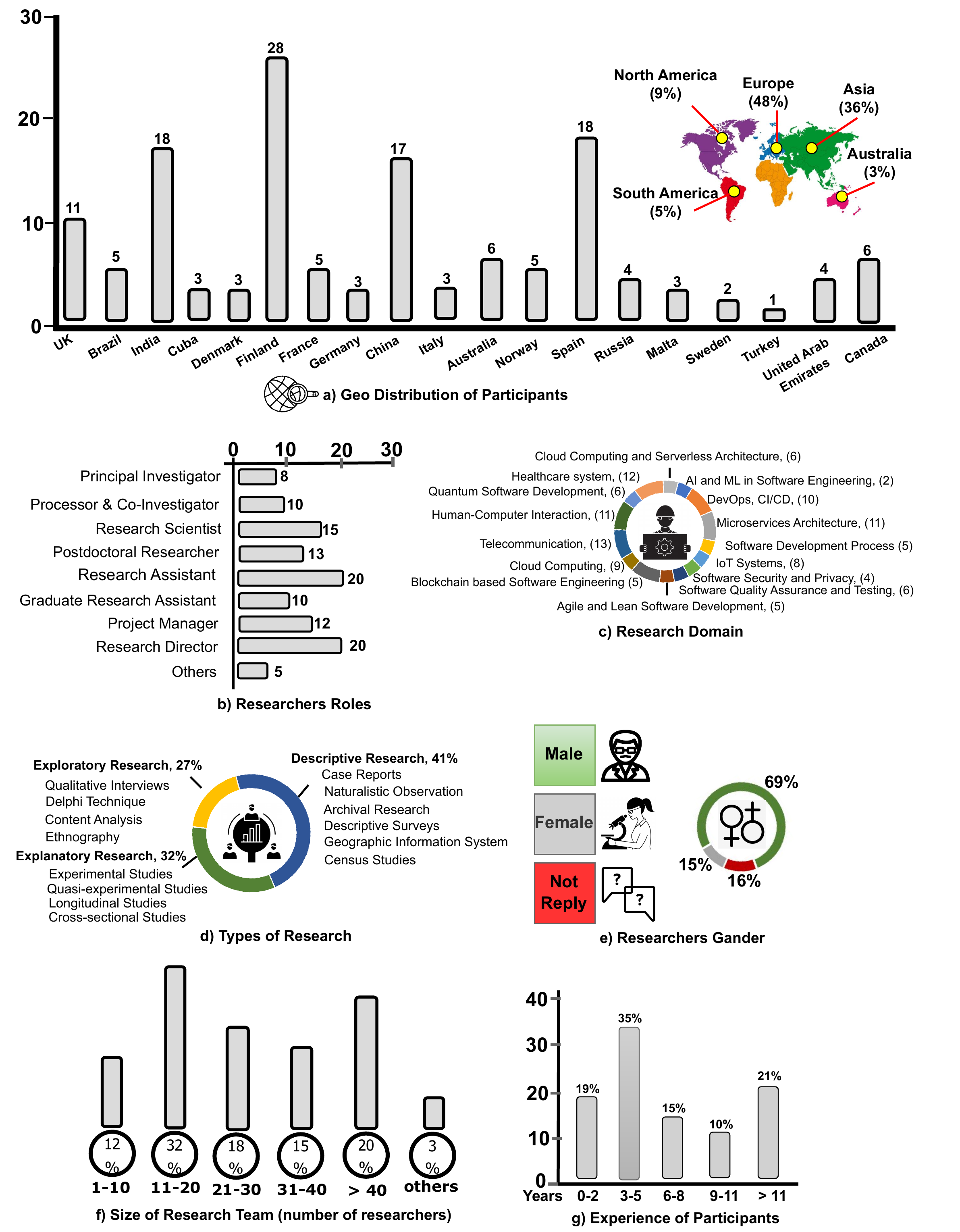}
\caption{Demographic Information Of Survey Participants}
\label{Fig:Demographic Information Of Survey Participants}
\end{figure*}

We conducted a frequency analysis to systematically organize the descriptive data, which is well-suited for examining a group of variables for numeric and ordinal data. Our study included 113 respondents from 19 countries across 5 continents, representing 9 professional roles, 15 distinct research domains, and 3 different types of research (see Figure \ref{Fig:Demographic Information Of Survey Participants}(a-d)).

Through applying thematic mapping, we categorized the respondents' roles into nine different categories (see Figure \ref{Fig:Demographic Information Of Survey Participants}(b)). The results indicate that 20\% of the respondents were primarily distributed between research assistants and research directors. Additionally, the participants' research teams are conceptually organized across 15 key research domains (see Figure \ref{Fig:Demographic Information Of Survey Participants}(c)). We found that 13\% of participants were engaged in telecommunications research, while 12\% worked within the healthcare context (see Figure \ref{Fig:Demographic Information Of Survey Participants}(c)).

Regarding demographic information, 69\% of the survey participants were male (see Figure \ref{Fig:Demographic Information Of Survey Participants}(e)). As measured by the number of researchers, the research team size predominantly ranges from 11 to 20, accounting for 32\% of total responses (see Figure \ref{Fig:Demographic Information Of Survey Participants}(f)). Among all respondents, the majority (35\%) reported having 3-5 years of research experience (see Figure \ref{Fig:Demographic Information Of Survey Participants}(g)).

\begin{figure*}[h!]
 \centering
    \includegraphics[width=1\linewidth]{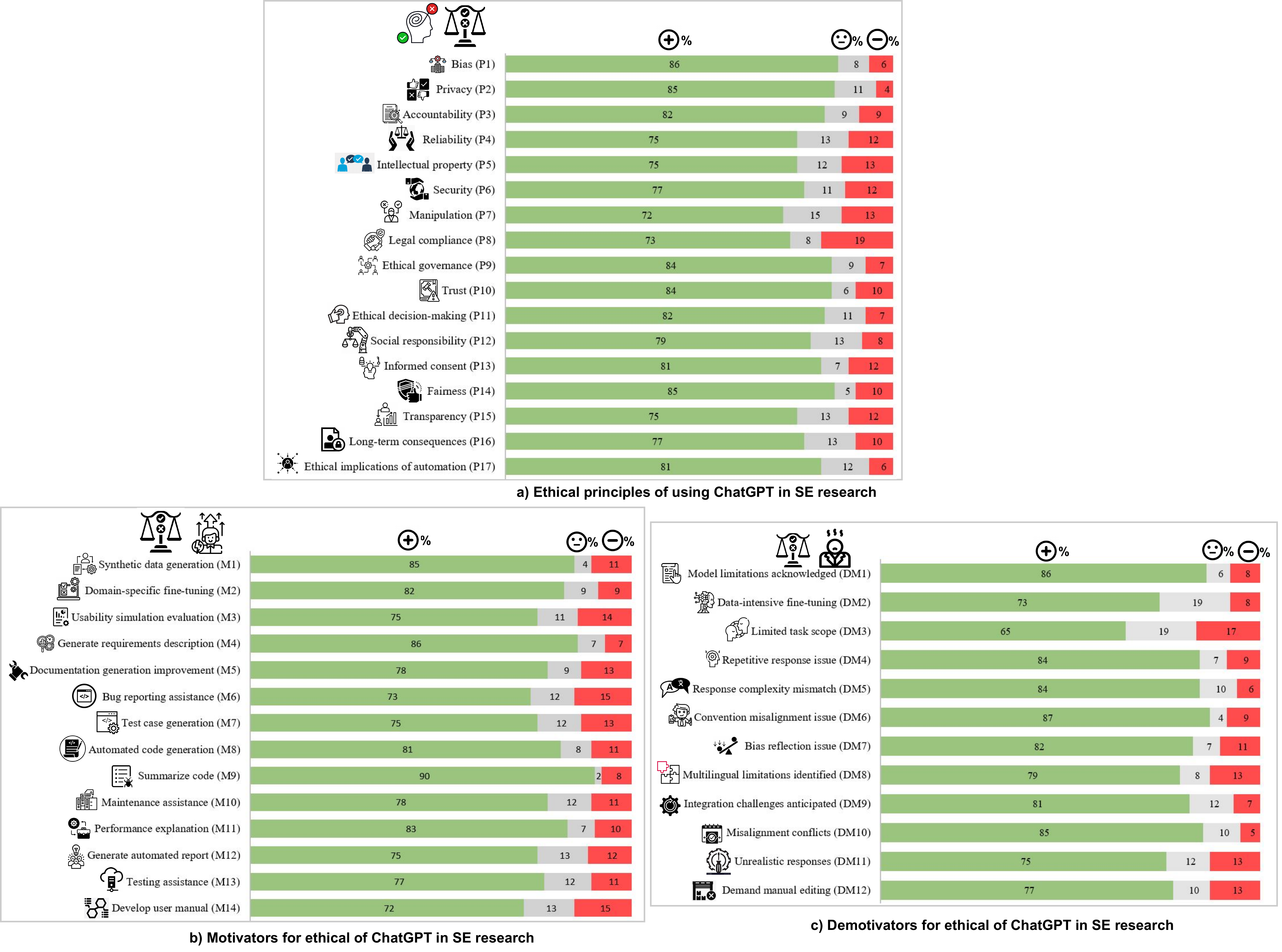}
\caption{Survey Participants Perceptions on Ethical Principles, relevant Motivators, and Demotivators}
\label{Fig:Survey Participants Perceptions on Ethical Principles, relevant Motivators, and Demotivators}
\end{figure*}
\subsubsection{Empirical Insights on Ethical Principles and related Motivators, and Demotivates}
The survey responses are classified as \textit{average agree, neutral} and \textit{average disagree} (see Figure \ref{Fig:Survey Participants Perceptions on Ethical Principles, relevant Motivators, and Demotivators}(a-c)). We observed that (approx. 80\%) of the respondents positively confirmed the findings of the literature survey, i.e.,  ethical principles of using ChatGPT in SE research, their relevant motivators, and demotivators.

The frequency analysis results show that a significant majority (86\%) of the survey participants consider \textit{P1 (Bias)} as an important ethical principle when using ChatGPT in SE research (Figure \ref{Fig:Survey Participants Perceptions on Ethical Principles, relevant Motivators, and Demotivators}(a)). The high percentage of participants who emphasize the importance of addressing bias in ChatGPT demonstrates a strong consensus among the SE research community. Addressing bias in AI language models like ChatGPT is crucial for ensuring reliable, valid, and generalizable SE research outcomes while promoting more inclusive, fair, and trustworthy AI tools \cite{ferrara2023should, lund2023chatgpt}. Furthermore, \textit{P2 (Privacy)} and \textit{P14 (Fairness)} are considered as the second most important (85\%) principles for ethical alignment of ChatGPT in SE research. \textit{Privacy}Privacy and \textit{Fairness} are considered vital to foster responsible, transparent, and accountable research while addressing the potential consequences of inadequate attention to data privacy and equitable treatment of individuals and groups \cite{mcgee2023ethics, rivas2023marketing}. 

\textit{P7 (Manipulation)} is considered the least significant principle (72\%) by the survey participants (Figure \ref{Fig:Survey Participants Perceptions on Ethical Principles, relevant Motivators, and Demotivators}(a)), potentially due to the nature of SE research and the context of ChatGPT usage. SE research focuses on software development and processes, which may make manipulation less relevant or critical compared to principles like bias, privacy, and fairness \cite{hacker2021manipulation, dwivedi2023so}. ChatGPT's usage in SE research may not involve as much end-user interaction as other AI applications, reducing perceived potential for manipulation. Despite this, researchers should remain vigilant and address any manipulative effects or unintended consequences associated with ChatGPT use \cite{tlili2023if, dwivedi2023so}. 

The survey results reveal that average 75\% of participants have confirmed the significance of the identified motivators to support the ethical principles of using  ChatGPT in SE research (see Figure \ref{Fig:Survey Participants Perceptions on Ethical Principles, relevant Motivators, and Demotivators}(b)). The most high frequency motivators - \textit{M9 (Summarize code, 90\%), M4 (generate requirements description, 86\%)}, and \textit{M1 (synthetic data generation, 85\%)} demonstrate the potential value and usefulness of ChatGPT in SE research when implemented responsibly. M9 streamlines code comprehension and maintenance, saving time and effort while maintaining code accuracy and efficiency \cite{liu2023summary}. M4 addresses the need for clear requirements descriptions, as ChatGPT can generate precise and coherent descriptions, reducing miscommunication and fostering better understanding among stakeholders \cite{hu2023opportunities}. M1 enhances the research process by providing realistic, anonymized data for extensive testing and validation without compromising privacy and ethics \cite{tang2023does}. These motivators emphasize the potential benefits of responsibly incorporating ChatGPT in SE research to improve various aspects of the software development life cycle.

Additionally, 80\% of participants on average believe that the identified demotivators could negatively impact the ethical principles of using ChatGPT in SE research (see Figure \ref{Fig:Survey Participants Perceptions on Ethical Principles, relevant Motivators, and Demotivators}(c)). Specifically, \textit{DM6 (Convention misalignment issue), DM1 (Model limitations acknowledged)}, and \textit{DM10 (Misalignment conflicts)} were considered significant by 87\%, 86\%, and 85\% of survey participants, respectively. Because DM6 emphasizes the need to address inconsistencies of ChatGPT's output in terms of coding conventions or best practices, which can hinder the development process and affect code quality \cite{frieder2023mathematical, mizumoto2023exploring}. DM1 emphasizes the importance of acknowledging ChatGPT's limitations, such as biases or inaccuracies, to prevent over-reliance on the model and maintain ethical research standards \cite{lecler2023revolutionizing}. DM10 stresses the need to resolve conflicts between ChatGPT's output and a project's goals, values, or ethical principles to maintain trust and collaboration among stakeholders \cite{pardos2023learning}. Addressing these demotivators is crucial for ensuring AI language models' (i.e., ChatGPT) ethical and responsible use in SE research.

\subsection{ISM Based Modeling of Ethical Principles} \label{Sec:ISMA Based Modeling of Ethical Principles}
By employing the ISM approach to categorize the identified ethical principles into five distinct levels, researchers and organizations can gain a deeper understanding of the relationships between these principles. This structure promotes more responsible, ethical, and socially conscious use of ChatGPT in SE research. The ISM analysis was conducted using the expert's opinions collected using a questionnaire survey (as discussed in Section \ref{Sec:ISM Approach}). The collected responses were summarized and used the following symbols to indicate the direction of the relationship between the ChatGPT principles (principle m and principle n).

\begin{itemize}
    \item “V" indicates the relationship of principle \textit{m} to \textit{n}. 
    \item “A" indicates the relationship of principle \textit{n} to \textit{m}.
    \item “X" when both principals \textit{m} and \textit{n} reach each other.
    \item “O" presents the situation when there is no relationship between principal \textit{m} and principal \textit{n}.
\end{itemize}

Using the experts' opinions and the above-discussed symbols, we developed the Structural Self-Interaction Matrix (SSIM) as presented in Table \ref{tab:: SSIM matrix}. The data shown in Table \ref{tab:: SSIM matrix} indicates no relationship between \textit{P1 (Bias) } and \textit{P17 (Ethical implications of automation)}, represented by an `O'. In contrast, a significant relationship is observed between \textit{P13 (Informed consent)} and \textit{P1 (Bias)}, where `V' signifies this relationship. Notably, we discovered no instances of `A' or `X' type relationships among the principles.

\begin{table*}[h!]
\centering
\caption{Structural Similarity Index Measure (SSIM) Matrix}
\label{tab:: SSIM matrix}
\begin{tabular}{|c|*{17}{c|}}
\hline
 & P17 & P16 & P15 & P14 & P13 & P12 & P11 & P10 & P9 & P8 & P7 & P6 & P5 & P4 & P3 & P2 & P1 \\
\hline
P1 & o & o & o & o & v & v & o & v & o & o & v & o & v & o & o & o &\cellcolor{lightgray}*\\
P2 & v & v & o & o & x & v & o & v & v & o & v & v & v & o & o & \cellcolor{lightgray}* &\cellcolor{lightgray} * \\
P3 & o & o & o & o & o & v & o & v & v & o & v & v & o & o & \cellcolor{lightgray}* &\cellcolor{lightgray} *& \cellcolor{lightgray} * \\
P4 & v & o & o & o & o & o & o & o & v & v & v & v & v &\cellcolor{lightgray} * &\cellcolor{lightgray} * &\cellcolor{lightgray} * &\cellcolor{lightgray} * \\
P5 & o & o & o & v & o & v & o & o & v & v & v & v &\cellcolor{lightgray} * &\cellcolor{lightgray} * &\cellcolor{lightgray} * &\cellcolor{lightgray} * &\cellcolor{lightgray} * \\
P6 & o & o & v & o & v & o & o & o & v & v & v &\cellcolor{lightgray} * &\cellcolor{lightgray} * &\cellcolor{lightgray} * &\cellcolor{lightgray} * &\cellcolor{lightgray} * &\cellcolor{lightgray} * \\
P7 & o & o & o & o & o & o & o & o & o & o &\cellcolor{lightgray} * &\cellcolor{lightgray} * &\cellcolor{lightgray} * &\cellcolor{lightgray} * &\cellcolor{lightgray} * &\cellcolor{lightgray} * &\cellcolor{lightgray} * \\
P8 & v & o & o & v & o & v & o & v & o &\cellcolor{lightgray} * &\cellcolor{lightgray} * &\cellcolor{lightgray} * &\cellcolor{lightgray} * &\cellcolor{lightgray} * &\cellcolor{lightgray} * &\cellcolor{lightgray} * &\cellcolor{lightgray} * \\
P9 & v & o & o & o & o & o & o & v &\cellcolor{lightgray} * &\cellcolor{lightgray} * &\cellcolor{lightgray} * &\cellcolor{lightgray} * &\cellcolor{lightgray} * &\cellcolor{lightgray} * &\cellcolor{lightgray} * &\cellcolor{lightgray} * &\cellcolor{lightgray} * \\
P10 & o & o & o & v & v & o & v &\cellcolor{lightgray} * &\cellcolor{lightgray} * &\cellcolor{lightgray} * &\cellcolor{lightgray} * &\cellcolor{lightgray} * &\cellcolor{lightgray} * &\cellcolor{lightgray} * &\cellcolor{lightgray} * &\cellcolor{lightgray} * &\cellcolor{lightgray} * \\
P11 & o & o & o & o & v & o &\cellcolor{lightgray} * &\cellcolor{lightgray} * &\cellcolor{lightgray} * &\cellcolor{lightgray} * &\cellcolor{lightgray} * &\cellcolor{lightgray} * &\cellcolor{lightgray} * &\cellcolor{lightgray} * &\cellcolor{lightgray} * &\cellcolor{lightgray} * &\cellcolor{lightgray} * \\
P12 & v & o & o & o & o &\cellcolor{lightgray} * &\cellcolor{lightgray} * &\cellcolor{lightgray} * &\cellcolor{lightgray} * &\cellcolor{lightgray} * &\cellcolor{lightgray} * &\cellcolor{lightgray} * &\cellcolor{lightgray} * &\cellcolor{lightgray} * &\cellcolor{lightgray} * &\cellcolor{lightgray} * &\cellcolor{lightgray} * \\
P13 & o & o & o & o &\cellcolor{lightgray} * &\cellcolor{lightgray} * &\cellcolor{lightgray} * &\cellcolor{lightgray} * &\cellcolor{lightgray} * &\cellcolor{lightgray} * &\cellcolor{lightgray} * &\cellcolor{lightgray} * &\cellcolor{lightgray} * &\cellcolor{lightgray} * &\cellcolor{lightgray} * &\cellcolor{lightgray} * &\cellcolor{lightgray} * \\
P14 & o & o & v &\cellcolor{lightgray} * &\cellcolor{lightgray}\cellcolor{lightgray}\cellcolor{lightgray}\cellcolor{lightgray}\cellcolor{lightgray}\cellcolor{lightgray}\cellcolor{lightgray}\cellcolor{lightgray}\cellcolor{lightgray}\cellcolor{lightgray} * &\cellcolor{lightgray}\cellcolor{lightgray}\cellcolor{lightgray}\cellcolor{lightgray}\cellcolor{lightgray}\cellcolor{lightgray}\cellcolor{lightgray}\cellcolor{lightgray}\cellcolor{lightgray} * &\cellcolor{lightgray}\cellcolor{lightgray}\cellcolor{lightgray}\cellcolor{lightgray}\cellcolor{lightgray}\cellcolor{lightgray}\cellcolor{lightgray}\cellcolor{lightgray} * &\cellcolor{lightgray}\cellcolor{lightgray}\cellcolor{lightgray}\cellcolor{lightgray}\cellcolor{lightgray}\cellcolor{lightgray}\cellcolor{lightgray} * &\cellcolor{lightgray}\cellcolor{lightgray}\cellcolor{lightgray}\cellcolor{lightgray}\cellcolor{lightgray}\cellcolor{lightgray} * &\cellcolor{lightgray}\cellcolor{lightgray}\cellcolor{lightgray}\cellcolor{lightgray}\cellcolor{lightgray} * &\cellcolor{lightgray}\cellcolor{lightgray}\cellcolor{lightgray}\cellcolor{lightgray} * &\cellcolor{lightgray}\cellcolor{lightgray}\cellcolor{lightgray} * &\cellcolor{lightgray}\cellcolor{lightgray} * &\cellcolor{lightgray} * &\cellcolor{lightgray}\cellcolor{lightgray} * &\cellcolor{lightgray} * &\cellcolor{lightgray} * \\
P15 & o & o &\cellcolor{lightgray}\cellcolor{lightgray} * &\cellcolor{lightgray} * &\cellcolor{lightgray} * &\cellcolor{lightgray}  * &\cellcolor{lightgray}  * &\cellcolor{lightgray}  * &\cellcolor{lightgray}  * &\cellcolor{lightgray}  * &\cellcolor{lightgray}  * &\cellcolor{lightgray}  * &\cellcolor{lightgray}  * &\cellcolor{lightgray}  * &\cellcolor{lightgray}  * &\cellcolor{lightgray}  * &\cellcolor{lightgray}* \\
P16 & o &\cellcolor{lightgray}  * &\cellcolor{lightgray}  * &\cellcolor{lightgray}  * &\cellcolor{lightgray}  * &\cellcolor{lightgray}  * &\cellcolor{lightgray}  * &\cellcolor{lightgray}  * &\cellcolor{lightgray}  * &\cellcolor{lightgray}  * &\cellcolor{lightgray}  * &\cellcolor{lightgray}  * &\cellcolor{lightgray}  * &\cellcolor{lightgray}  * &\cellcolor{lightgray}  * &\cellcolor{lightgray}  * &\cellcolor{lightgray} * \\
P17 &\cellcolor{lightgray}* &\cellcolor{lightgray}* &\cellcolor{lightgray}* &\cellcolor{lightgray}* &\cellcolor{lightgray}* &\cellcolor{lightgray}* &\cellcolor{lightgray}* &\cellcolor{lightgray}* &\cellcolor{lightgray}* &\cellcolor{lightgray}* &\cellcolor{lightgray}* &\cellcolor{lightgray}* &\cellcolor{lightgray}* &\cellcolor{lightgray}* &\cellcolor{lightgray}* &\cellcolor{lightgray}* &\cellcolor{lightgray}* \\
\hline
\end{tabular}
\end{table*}

In the next step of ISM analysis, the transformation of the SSIM data into a binary reachability matrix is performed. This process entails converting values represented by `V', `A', `S', and `O' into binary digits (0, 1) using the following conversion rules. For instance, If the value of \textit{m} and \textit{n} in SSIM is V, then we replace it with 1; else, the assigned value is 0. If the value of \textit{m} and \textit{n} in SSIM is A, then it is replaced with 0; else it becomes 1. If the value of \textit{m} and \textit{n} in SSIM is X, then it is replaced with 1; and give 1 to \textit{n} and \textit{m} entry. If the value of \textit{m} and \textit{n} in SSIM is O, then it will replace with 0; and for \textit{m} and \textit{n}, the assigned value is also 0. These rules are designed to ensure that the binary reachability matrix accurately reflects the relationships outlined in the SSIM. Once this transformation is completed, we then enhance the reachability matrix by applying a transitivity check, as discussed in Section \ref{Sec:ISM Approach}. This introduces a `1*' value to capture transitivity, helping to fill potential gaps in the expert data collected during the SSIM development stage. The application of this transitivity check is further detailed in Table \ref{tab: Reachability Matrix}. This rigorous process ensures our ISM analysis is robust, comprehensive, and reflective of the expert insights gathered.

\begin{table*}[h!]
\centering
\caption{Reachability Matrix }
\label{tab: Reachability Matrix}
\begin{tabular}{|c|*{19}{c|}}
\hline
    & P1 & P2 & P3 & P4 & P5 & P6 & P7 & P8 & P9 & P10 & P11 & P12 & P13 & P14 & P15 & P16 & P17 & DIV & RANK \\
\hline
P1 & \cellcolor{lightgray}1 & 0 & 0 & 0 & 1 & 0 & 1 & 0 & 0 & 1 & 0 & 1 & 1 & 0 & 0 & 0 & 0 & 6 & 4 \\
P2 & 0 & \cellcolor{lightgray}1 & 0 & 0 & 1 & 0 & 1 & 0 & 1 & 1 & 0 & 1 & 0 & 0 & 0 & 1 & 1 & 8 & 6 \\
P3 & 0 & 0 & \cellcolor{lightgray}1 & 0 & 0 & 0 & 0 & 0 & 0 & 1 & 1 & 0 & 1 & 0 & 0 & 0 & 0 & 4 & 2 \\
P4 & 1 & 1 & 0 & \cellcolor{lightgray}1 & 1 & 1 & 1 & 1 & 1 & 0 & 0 & 0 & 0 & 1* & 0 & 0 & 1 & 9 & 8 \\
P5 & 1 & 0 & 0 & 1* & \cellcolor{lightgray}1 & 1 & 1 & 1 & 1 & 0 & 0 & 1 & 0 & 1 & 0 & 0 & 0 & 8 & 7 \\
P6 & 0 & 1 & 0 & 1 & 0 & \cellcolor{lightgray}1 & 1 & 1 & 1 & 0 & 0 & 1 & 0 & 1 & 0 & 0 & 0 & 8 & 6 \\
P7 & 1 & 0 & 0 & 0 & 0 & 1* & \cellcolor{lightgray}1 & 1* & 1* & 0 & 0 & 0 & 0 & 0 & 0 & 0 & 0 & 2 & 3 \\
P8 & 0 & 0 & 0 & 0 & 0 & 1 & 0 & \cellcolor{lightgray}1 & 0 & 1 & 0 & 1 & 0 & 1 & 0 & 0 & 1 & 6 & 4 \\
P9 & 1 & 0 & 0 & 1 & 0 & 1 & 0 & 0 & \cellcolor{lightgray}1 & 1 & 0 & 0 & 0 & 0 & 1* & 0 & 1 & 6 & 5 \\
P10 & 1 & 1 & 0 & 1 & 1 & 0 & 0 & 0 & 0 & \cellcolor{lightgray}1 & 1 & 0 & 1 & 1 & 0 & 0 & 0 & 8 & 6 \\
P11 & 1 & 0 & 1 & 0 & 0 & 0 & 0 & 0 & 0 & 0 & \cellcolor{lightgray}1 & 0 & 1 & 0 & 0 & 0 & 0 & 4 & 2 \\
P12 & 1 & 0 & 1 & 0 & 1 & 1 & 0 & 0 & 0 & 0 & 0 & \cellcolor{lightgray}1 & 0 & 0 & 0 & 0 & 1 & 6 & 4 \\
P13 & 0 & 0 & 0 & 0 & 0 & 0 & 0 & 0 & 0 & 0 & 1 & 0 & \cellcolor{lightgray}1 & 0 & 0 & 0 & 0 & 2 & 1 \\
P14 & 0 & 0 & 0 & 0 & 0 & 0 & 0 & 1* & 1 & 1 & 0 & 0 & 0 & \cellcolor{lightgray}1 & 1 & 0 & 0 & 4 & 3 \\
P15 & 1 & 1* & 0 & 1 & 0 & 1 & 0 & 1 & 0 & 0 & 0 & 0 & 0 & 0 & \cellcolor{lightgray}1 & 0 & 0 & 5 & 4 \\
P16 & 0 & 0 & 0 & 0 & 1 & 1 & 1 & 1 & 1 & 0 & 0 & 0 & 0 & 0 & 0 & \cellcolor{lightgray}1 & 0 & 6 & 4 \\
P17 & 0 & 0 & 0 & 0 & 0 & 1 & 1 & 1* & 1 & 1 & 0 & 0 & 0 & 0 & 0 & 0 & \cellcolor{lightgray}1 & 5 & 4 \\
DIV & 9 & 4 & 3 & 5 & 7 & 9 & 8 & 6 & 8 & 8 & 4 & 6 & 5 & 5 & 2 & 2 & 6 & & \\
RANK & 8 & 4 & 2 & 5 & 6 & 2 & 7 & 8 & 8 & 7 & 3 & 5 & 4 & 5 & 2 & 1 & 5 & & \\
\hline
\end{tabular}
\end{table*}

\begin{figure}[htbp]
\centering
\includegraphics[width=\linewidth]{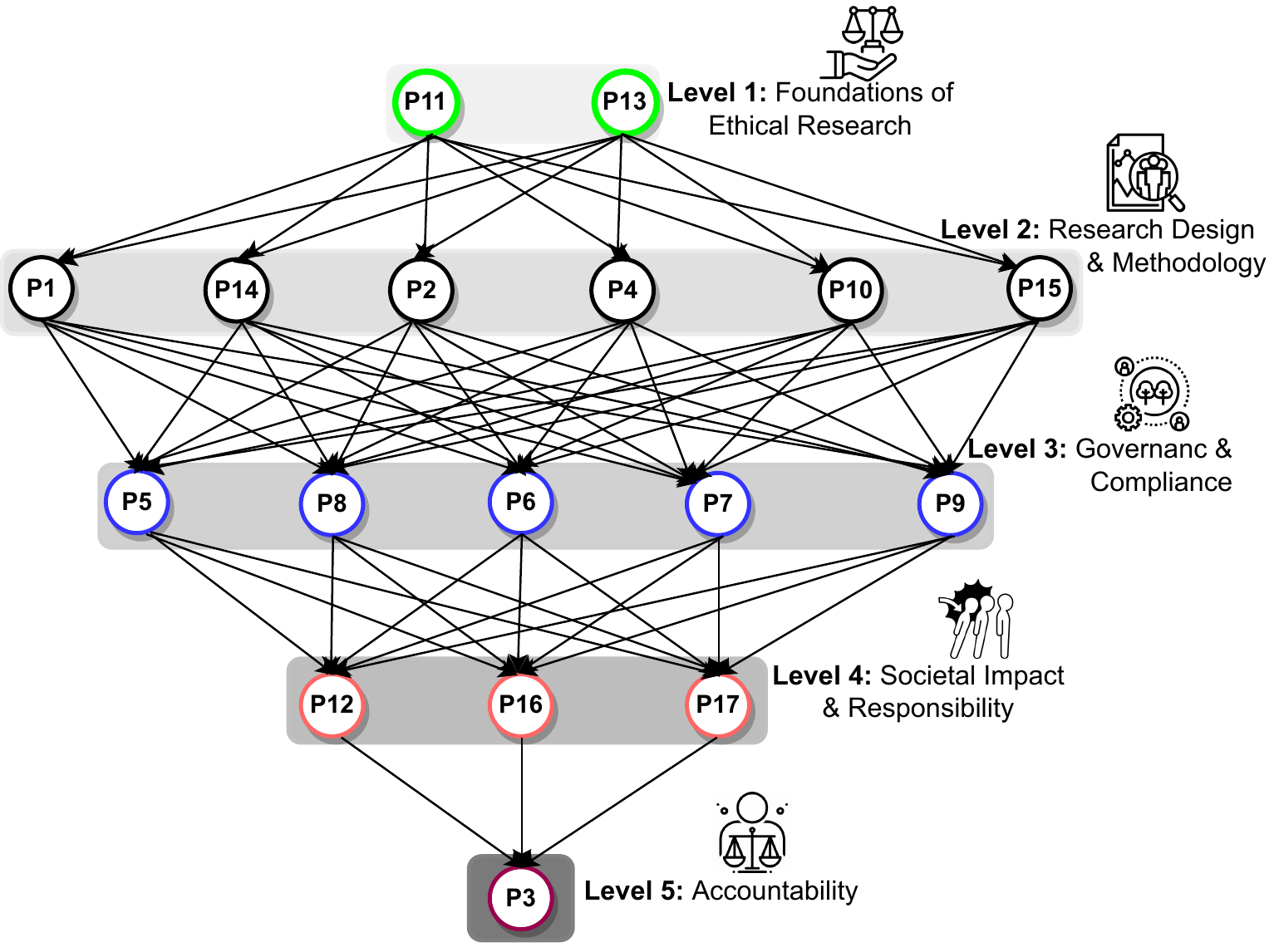}
\caption{Leveling of Ethics Principles using ChatGPT}
\label{fig:Leveling of Ethics Principles using ChatGPT}
\end{figure}

Warfield \cite{warfield1974developing} stated that the reachability set consists of the variables itself and other variables, which it may help to achieve, whereas the antecedent set consists of the variable itself and other variables, which may help achieving it. Thereafter, the intersection of these sets is derived for all the variables. Subsequently, the intersection of these sets is calculated for all variables. Variables that have the same sets of reachability and intersection are placed at the top level in the ISM hierarchy. These top-tier variables do not contribute to achieving any other variables above their level. Once this apex element is identified, it is isolated from the rest. This methodology is repeated to pinpoint the variables for each subsequent level. This process is continued until every variable’s level is determined. These levels are instrumental in constructing the digraph and the ISM model.
In our study, we have identified seventeen ethical principles for using ChatGPT in SE research. The finalized levels, as determined by our analysis, are illustrated in Figure \ref{fig:Leveling of Ethics Principles using ChatGPT}. A comprehensive analysis is provided at the following link\footnote{\url{https://tinyurl.com/39k3xhex}}.

Using large language model systems like ChatGPT requires a multi-level approach to ensure ethical research and responsible innovation. \textit{Level 1 (Foundations of Ethical Research)} focuses on \textit{P11 (Ethical decision-making)} and \textit{P13 (Informed consent)}, safeguarding the rights and dignity of participants. \textit{Level 2 (Research Design \& Methodology)} addresses \textit{P1 (Bias), P14 (Fairness), P2 (Privacy), P4 (Reliability), P10 (Trust)} and \textit{P15 (Transparency)} aligning AI systems with societal values and user expectations. \textit{Level 3 (Governance \& Compliance)} emphasizes adherence to organizational policies and legal frameworks, including \textit{P5 (Intellectual property), P6 (Security)} and \textit{P9 (Ethical governance)}. \textit{Level 4 (Societal Impact \& Responsibility)} highlights the importance of considering broader societal implications, such as misinformation, biased outcomes, and potentially harmful applications. Lastly, \textit{Level 5 (Accountability) } underscores the significance of fostering a collaborative environment and shared responsibility among researchers, organizations, and AI systems, bolstering stakeholder \textit{P15 (Transparency)} and \textit{P10 (Trust)}. By addressing these levels, researchers can ensure responsible development, deployment, and use of ChatGPT while upholding ethical standards and safeguarding stakeholder interests.

\subsection{MICMAC analysis}
MICMAC is abbreviated by matrix cross-impact matrix multiplication applied to the classification of identified principles. The MICMAC analysis assists in examining the key principles that drive the ethical aspect of using ChatGPT in SE research. According to Attri et al. \cite{attri2013analysis}, the MICMAC approach “\textit{is an analysis to examine principles drive power and dependence power}". The principles are classified into four clusters based on their driving and dependence power.

\begin{itemize}
    \item Autonomous cluster: the principles belonging to this cluster have weak driving and dependence power. They are mostly disconnected from the ethical scope due to weak links. Hence, these principles have a weak influence on the whole system (other principles). 
    \item Linkage cluster: the principles belonging to this cluster have strong driving and dependence power and affect other principles due to strong linkage. 
    \item Dependent cluster: the principles belonging to this cluster have strong dependence power but have weak driving power.
    \item Independent cluster: the principles belonging to this cluster have weak, dependent power but have strong driving power; they are also known as key principles. 
\end{itemize}

\begin{figure}[htbp]
\centering
\includegraphics[width=\linewidth]{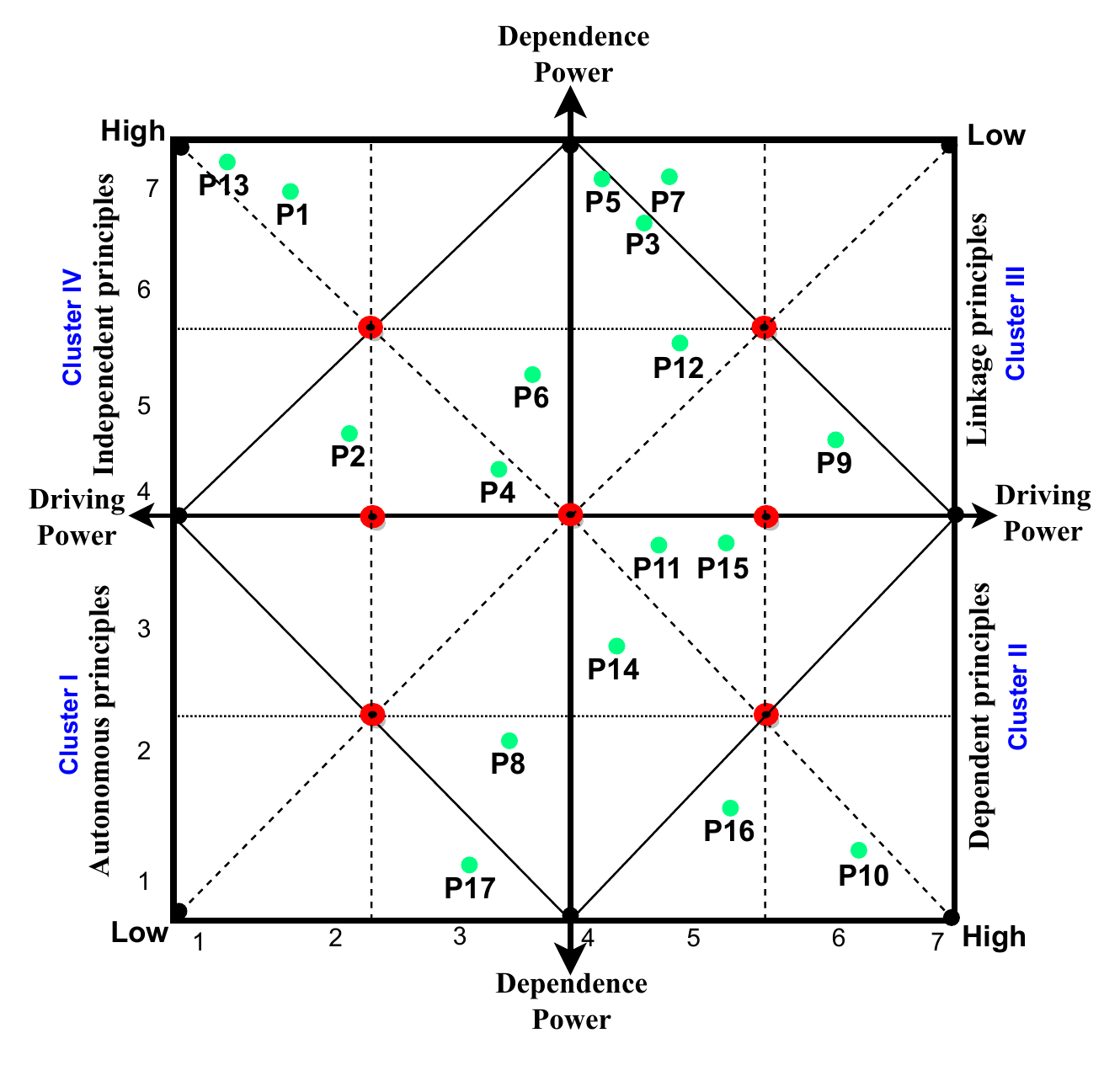}
\caption{Graphical view of the MICMAC analysis}
\label{fig:Graphical view of MICMAC analysis}
\end{figure}

After developing the hierarchical ISM model for ethical principles of ChatGPT using the ISM analysis, we conducted a MICMAC analysis based on the conical matrix provided at the given link\footnote{\url{https://tinyurl.com/39k3xhex}}. We employed the classification approach proposed by Kannan et al.~\cite{kannan2009hybrid} for the MICMAC-based categorization and present the results in Figure \ref{fig:Graphical view of MICMAC analysis}. We identified and organized the ethical principles for conducting SE research using ChatGPT into four distinct clusters, as determined by the MICMAC analysis.

First, the independent cluster includes principles, such as addressing \textit{P1 (Bias), ensuring P2 (Privacy), maintaining P4 (Reliability), implementing P6 (Security)}, and obtaining \textit{P13 (Informed consent)}. These principles form the foundation of ethical research practice and influence other aspects of ChatGPT development, characterized by weak dependence power and strong driving power. Second, the dependent principles, such as \textit{P10 (Trust), P11 (Ethical decision-making), P14 (Fairness), P15 (Transparency)}, and \textit{P16 (Long-term consequences)} signify the outcomes of ethical research. These principles serve as indicators of effective practice and success measurement in SE research using ChatGPT, having strong dependence power but weak driving power.

Third, the linkage variables connect independent and dependent principles, including \textit{P3 (Accountability), P5 (Intellectual property), P7 (manipulation), P9 (Ethical governance)} and \textit{P12 (Social responsibility)}. This cluster ensures a comprehensive and ethically sound approach to ChatGPT development, with principles exhibiting strong driving and dependence power due to their robust linkage. Finally, autonomous variables, like \textit{P8 (Legal compliance)} and \textit{P17 (Ethical implications of automation)} hold unique positions within the analysis. These principles play crucial roles in understanding the broader implications of ChatGPT development and fostering a responsible, sustainable AI ecosystem that aligns with societal expectations and legal requirements. However, with weak driving and dependence power, these principles are mostly disconnected from the ethical scope and have a minor impact on the overall system. 

In conclusion, our MICMAC analysis evaluates the relationships between ethical principles, taking into account their driving and dependence power. This insight enables the research community to understand the varying dynamics of these principles, such as those with driving power over others, those that are independent yet influential, those that are completely autonomous without any driving or dependence power, and those wholly dependent on other principles. By recognizing these distinct characteristics, researchers can devise more effective strategies for the ethical adoption of ChatGPT in SE research.

%% file: Study-Implications.tex
\section{Study Implications} \label{Sec: Study Implications}
The findings of this study can be leveraged by the research community in order to establish best practices to adopt ethically align ChatGPT within the realm of SE research. Here are the key implications derived from the study's findings:
\begin{itemize}
    \item \textbf{Improved Collaboration and Efficiency:} Software engineering researchers can utilize the results of this study to effectively incorporate ChatGPT into their research processes. The identified motivators, such as \textit{M9 (Summarizing code)} and \textit{M4 (Generating requirements descriptions)}, can lead to better knowledge extraction and seamless communication among developers, researchers, and AI-assisted code generation systems, ultimately enhancing efficiency and collaboration within the SE research community.
    \item \textbf{Ethical Research Practices:} The identification of 17 ethical principles and their corresponding motivators and demotivators can guide SE researchers in conducting ethically responsible research using ChatGPT. By considering critical principles such as \textit{P1 (Bias), P2 (Privacy)} and \textit{P14 (Fairness)}, researchers can ensure that their work with ChatGPT adheres to the highest ethical standards, minimizing potential risks and harm.
    \item \textbf{Informed Decision-Making:} The level-based decision model (developed using ISM) and cluster-based decision model (developed using MICMAC) provide researchers with valuable insights into the relationships among various ethical principles. Understanding the dependence and driving power of these principles can help researchers make more informed decisions about the use of ChatGPT in their work, ensuring that they address key ethical considerations throughout the research process.
    \item \textbf{Addressing Ethical Challenges:} The study highlights specific demotivators that can negatively impact the consideration of ethical principles in SE research. By acknowledging and addressing these demotivators, such as \textit{DM6 (Convention misalignment issues)} and \textit{DM1 (Model limitations)}, researchers can develop strategies to mitigate potential ethical concerns, leading to more responsible and conscientious use of ChatGPT in their research.
    \item \textbf{Team Development:} Organizations can use the identified motivators and demotivators, as well as the ethical principles outlined in this study, as a guide to selecting and training researchers and AI professionals for ChatGPT implementation in SE research. These factors can serve as a risk mitigation strategy, ensuring the team has the necessary skills to effectively and ethically utilize ChatGPT in SE research projects while addressing potential challenges and ethical concerns.
    \item \textbf{Research Process Assessment:} The study findings offer researchers valuable insights into the ethical considerations of ChatGPT in SE research projects. By understanding the identified ethical principles, motivators, and demotivators, SE researchers can better assess their current projects' strengths and weaknesses in relation to ChatGPT implementation. By leveraging these insights, improvements to ChatGPT usage within SE research processes can be more effectively planned, managed, and targeted. Ultimately, this comprehensive understanding of ChatGPT's ethical and practical implications will enable organizations to be better positioned to conduct responsible, efficient, and collaborative SE research with the assistance of ChatGPT.  
\end{itemize}

%% file: Threats-to-Validity.tex
\section{Threats to Validity} \label{Sec:Threats to Validity}
Several threats could potentially affect the validity of the findings of this study. Accordingly, we identified and categorized potential threats, aligning them with internal validity, external validity, construct validity, and conclusion validity as per the guidelines defined by Easterbrook et al. \cite{easterbrook2008selecting}.

\subsection{Internal Validity} 
Internal validity is the degree to which the results of observation — namely, the causal relationships — are trustworthy and not influenced by other factors or biases. The potential internal validity threat in this study is the understandability and interpretation of the survey content. The survey respondents may have a different understanding of the survey questions, which could bias the responses. To mitigate this threat, we piloted the instrument, seeking feedback from SE researchers to enhance the clarity and readability of the survey content prior to its final distribution. 

\subsection{External Validity}
External validity is the extent to which the results of a study can be generalized or applied to other situations, populations, or settings. In this study, the questionnaire data were collected from 113 researchers, which may not be representative of the broader SE research community. This could limit the generalizability of the findings. Nonetheless, we gathered 113 valid responses from 19 countries across five different continents. The survey participants had a diverse range of experience, fulfilled various roles in different projects, and worked in research teams of differing sizes (see Figure \ref{Fig:Demographic Information Of Survey Participants}). We agree that the study findings could not be generalized to a larger scale; however, based on the details demographics of the survey participants, the overall results could be generalized to some extent.

\subsection{Construct Validity}
Construct validity refers to the degree to which a test or experiment measures what it claims to be measuring. In this study, the constructs such as "motivators," "demotivators," and "ethical principles" may not have been defined clearly enough, leading to potential misinterpretation. However, we mitigated this threat by defining and elaborating on the mentioned constructs based on the literature survey. The identified “motivators", “demotivators", and “ethical principles" are comprehensively discussed in Section \ref{Sec: literature Survey}. Moreover, the survey questionnaire was piloted based on the expert's opinion to improve the interpretations of the survey variables (constructs).

\subsection{Conclusion Validity}
Conclusion validity is concerned with the relationship between the treatment and the outcome and whether any observed effect in the data is real or not. One possible threat to the conclusion validity is that with only 113 respondents, the statistical power may be insufficient to detect meaningful differences or relationships. However, based on the existing relevant studies and the novelty of the research field, the given sample size is strong enough to draw the study's conclusions. Moreover, we plan to extend this study by widening the pool of potential respondents, extending the data collection period, and using different methods to reach the potential population (see Section \ref{sec:Future Plans}). Finally, all the authors were invited to participate in the brainstorming sessions to collaboratively dissect the primary findings and formulate definitive conclusions. 

%% file: Conclusions-Future-Plan.tex
\section{Conclusions and Future Plans}\label{Sec:Conclusions and Future Plans}
We will now present a summary of the conclusions drawn from the study findings, along with a detailed roadmap outlining potential avenues for future exploration.

\subsection{Conclusions}\label{Sec:Conclusions}
ChatGPT enhances efficiency in knowledge extraction and collaboration within SE research. Its capacity to produce realistic and contextually appropriate language renders it an attractive tool for use in this research field. However, ethical concerns such as plagiarism, privacy, data security, and the risk of generating biased or harmful data must be addressed. This study explores the motivators, demotivators, and ethical principles associated with using ChatGPT in SE research.

We conducted a literature survey, identified 17 ethical principles and their corresponding 14 motivators and 12 demotivators for using ChatGPT in SE research, as detailed in Section \ref{Sec:Literature Survey Findings}. These motivators and demotivators were subsequently mapped to the 17 identified principles. The principles highlight crucial areas that the SE research community must consider in order to conduct ethically responsible research. The associated motivators represent factors that can support adherence to these principles. Conversely, demotivators are factors that may obstruct the consideration of ethical principles when using ChatGPT in SE research.

To empirically evaluate the significance of the identified principles and their associated motivators and demotivators, we conducted a questionnaire survey involving SE researchers. The frequency analysis highlights a strong consensus among survey participants, with 86\% identifying \textit{P1 (Bias)} P1 (Bias) as a critical ethical principle in using ChatGPT for SE research. Furthermore, 85\% deemed \textit{P2 (Privacy)}  and \textit{P14 (Fairness)} as equally important principles. Interestingly, \textit{P7 (manipulation)} was considered less significant at 72\%. The results also demonstrated that an average of 75\% of participants acknowledged the importance of specific motivators, such as \textit{M9 (Summarizing code), M4 (Generating requirements descriptions)}, and \textit{M1 (Synthetic data generation)}, in supporting ethical principles. Meanwhile, an average of 80\% agreed that identified demotivators, including \textit{DM6 (Convention misalignment issues), DM1 (Model limitations acknowledged)}, and \textit{DM10 (Misalignment conflicts)}, could negatively impact ethical principles. Addressing these concerns is important for ensuring the responsible use of ChatGPT in SE research.

We used the Interpretive Structural Modeling (ISM) approach to create level-based decision models and perform the MICMAC analysis for the development of cluster-based decision models. The ISM based decision model comprises five of levels: \textit{Level 1 (Foundations of Ethical Research)}, \textit{Level 2 (Research Design \& Methodology)}, \textit{Level 3 (Governance \& Compliance)}, \textit{Level 4 (Societal Impact \& Responsibility)}, and \textit{Level 5 (Accountability)}. These levels address various aspects, from ethical decision-making and informed consent to broader societal implications and shared responsibility among stakeholders. This level-based decision model illustrates the relationships among different principles, guiding the research community in addressing the ethical principles of using ChatGPT in SE research while considering the dependence and driving power of these principles on others.

The MICMAC analysis evaluated the relationships between ethical principles, categorizing them into four distinct clusters: independent, dependent, linkage, and autonomous. The MICMAC based decision model assists the SE research community in comprehending the diverse dynamics of the identified principles, including those that exert driving power over others, those that are independent yet carry significant influence, those that are completely autonomous devoid of any driving or dependence power, and those entirely reliant on other principles. By acknowledging these unique characteristics, researchers can formulate more effective strategies for the ethical implementation of ChatGPT in SE research.

\subsection{Future Plans}
\label{sec:Future Plans}

The ultimate aim of this research project is the development of comprehensive guidelines for using ChatGPT in SE research. This study presents a taxonomy (Figure \ref{Fig.Mapping of motivators and demotivators against research ethics}), developed based on the preliminarily investigated motivators, demotivators, and their possible impact on ChatGPT ethics aspects. The next steps are conducting a comprehensive multivocal literature review and questionnaire survey study to refine and identify the additional motivators and demotivators across ethical aspects of using ChatGPT in SE research. Below are the steps we will follow for the development of guidelines: 1) Perform an extensive multivocal literature review to identify a broader range of motivators, demotivators, and their potential impact on ethical aspects. 2) Design a survey questionnaire targeting SE researchers to validate the findings of multivocal literature review and explore additional motivators, demotivators, and ethical aspects of using ChatGpt in SE research. 3) Comparatively analyse the the findings of multivocal literature review and questionnaire survey. Identify common themes, trends, and patterns emerging from these data sources to create a taxonomy of motivators, demotivators, and their relationship to the ethical aspects. 4) Draft a set of guidelines based on the revised taxonomy addressing the ethical concerns of using ChatGPT in SE research. The guidelines should consider the motivators, demotivators, and their possible impact on ethical principles. 5) Invite AI ethics, software engineering, and ChatGPT application experts to review the draft guidelines. Collect their feedback and insights to refine and validate the guidelines. Incorporate their feedback to ensure the guidelines are robust, relevant, and effective in addressing ethical concerns. 6) Conduct academic case studies to evaluate the set of guidelines SE research and incorporate the potential changes based on the case studies' findings.
 7) Disseminate the final guidelines with the SE research community through conferences, workshops, and academic journals. Engage with the community to promote adopting and using the guidelines in practice. 8) Regularly review and update the guidelines in light of new developments in ChatGPT technology, ethical concerns, and feedback from the research community. Ensure these guidelines remain relevant and effective in addressing the ethical challenges of using ChatGPT in SE research.